\documentclass[prb,aps,twocolumn]{revtex4}

\usepackage{epsfig}
\usepackage{amsfonts}

\begin{document}

\title{NMR study of the spin excitations in the frustrated antiferromagnet Yb(BaBO$_3$)$_3$ with a triangular lattice}

\author{K. Y. Zeng$^{1}$}
\author{Long Ma$^{1}$}
\email{malong@hmfl.ac.cn}
\author{Y. X. Gao$^{2}$}
\author{Z. M. Tian$^{2}$}
\email{tianzhaoming@hust.edu.cn}
\author{L. S. Ling$^{1}$}
\author{Li Pi$^{1,3}$}
\email{pili@ustc.edu.cn}

\affiliation{$^{1}$Anhui Province Key Laboratory of Condensed Matter Physics at Extreme
Conditions, High Magnetic Field Laboratory, Chinese Academy of Sciences, Hefei
230031, China\\
$^{2}$ School of Physics and Wuhan National High Magnetic Field Center, Huazhong University of Science and Technology, Wuhan 430074, PR China\\
$^{3}$ Hefei National Laboratory for Physical Sciences at the Microscale, University of Science and Technology of China, Hefei 230026, China}

\date{\today}

\begin{abstract}

  In this paper, we study the spin excitation properties of the frustrated triangular-lattice antiferromagnet Yb(BaBO$_3$)$_3$ with nuclear magnetic resonance. From the spectral analysis, neither magnetic ordering nor spin freezing is observed with temperature down to $T=0.26$ K, far below its Curie-Weiss temperature $|\theta_w|\sim2.3$ K. From the nuclear relaxation measurement, precise temperature-independent spin-lattice relaxation rates are observed at low temperatures under a weak magnetic field, indicating the gapless spin excitations. Further increasing the field intensity, we observe a spin excitation gap with the gap size proportional to the field intensity. These phenomena suggest a very unusual strongly correlated quantum disordered phase, and the implications for the quantum spin liquid state are further discussed.

\end{abstract}

\maketitle

The persistent quantum fluctuations in geometrically frustrated antiferromagnets always lead to novel quantum ground states as well as exotic spin excitations. One of the archetypical case is the quantum spin liquid (QSL) state, where spins are highly entangled, strongly fluctuate but never order or freeze even at zero temperature\cite{Anderson_MRB_1973}. Various novel quantum properties such as fractional spin excitations, topological order, et al. can be expected in the QSLs\cite{Wen_RMP_2017}. Besides, the research on the QSLs is believed to be of significant importance for solving the puzzle of high-$T_c$ superconductivity\cite{Anderson_science_1987,Lee_RMP_78_17} as well as the quantum communication\cite{Jiang_NP_8_902}. Antiferromagnetically coupled small spins on the triangular or kagome lattice have supply people with promising route to realizing the QSL state.

Several geometrically frustrated antiferromagnets are argued to be the promising QSL candidates, including the hot materials of ZnCu$_3$(OH)$_6$Cl$_2$\cite{Shores_JACS_127_13462,Han_Nature_492_406,Fu_Science_350_655} and Cu$_3$Zn(OH)$_6$FBr\cite{Feng_CPL_34_077502} with kagome structure, and $\kappa-$(BEDT-TTF)$_2$Cu$_2$(CN)$_3$\cite{Shimizu_PRL_91_107001,Yamashita_NatPhys_4_459,Yamashita_NatPhys_5_44} and EtMe$_3$Sb[Pd(dmit)$_2$]$_2$\cite{Itou_PRB_77_104413,Yamashita_Science_328_1246,Yamashita_NC_2_275} with the triangular lattice. As analogs of the half spins carried by Cu$^{2+}$, the rare earth ions Yb$^{3+}$ located on triangular lattice, which can be treated as Kramers doublets with an effective spin of $J_{eff}=1/2$, supply an alternative way to approach the QSL state.

The newly discovered Yb$^{3+}$-based material YbMgGaO$_4$ is proposed to host a gapless U(1) QSL state as evidenced by the strong spin excitations with the temperature far below its Curie-Weiss temperature\cite{Li_PRL_115_167203, Li_PRL__117_097201} and the power-law temperature dependence of the specific heat\cite{Li_SR__5_16419}. However, the ground state is still highly controversial. The observed continuous spin excitation spectrum is first assigned to be spinons, and treated as a strong evidence for the QSL state\cite{Shen_Nature__540_559}. While the valence bond excitation is further proposed to be responsible for this continuum\cite{Li_PRL_122_137201,Kimchi_PRX_98_220409}.
From the thermal conductivity of YbMgGaO$_4$\cite{Xu_PRL__117_267202}, no significant magnetic excitation contribution is observed. The obvious broadening of the spin wave excitation in the polarized state, and broad crystal electric field excitation of Yb$^{3+}$ further indicate the strong disorder effect resulting from the random mixing between Mg$^{2+}$and Ga$^{3+}$\cite{Paddison_NP_13_117, Li_PRL_118_107202}. The frequency-dependent peak around $T=0.1$ K seen from $ac$-susceptibility, indicative of the spin-glass ground state\cite{Ma_PRL__120_087201}, further questions the existence of QSL state in YbMgGaO$_4$ and its counterpart YbZnGaO$_4$ with similar structure and properties. How severe influence on the ground state by the disorder effect is still unclear. A recent theoretical study shows that the disorder-free YbMgGaO$_4$ should exhibit a robust collinear magnetic order. It is the mixed Mg$^{2+}$/Zn$^{2+}$ and Ga$^{3+}$ site disorder effect that results in the spin disorder and give birth to the spin-liquid like behavior\cite{Zhu_PRL_119_157201}.

To clarify the controversial results about the ground state in YbMgGaO$_4$, studying new materials containing Yb$^{3+}$-triangular lattice is an alternative way with significant importance. In this paper, we employ nuclear magnetic resonance (NMR) as a local probe to perform detailed study on the static magnetism and spin dynamics in the triangular lattice antiferromagnet Yb(BaBO$_{3}$)$_{3}$\cite{Gao_JAC_745_396}. The Curie-Weiss fit to the low temperature $dc$-susceptibility gives a value of $\theta_w\sim-2.3$ K\cite{Gao_JAC_745_396}, comparable to that in YbZnGaO$_4$\cite{Ma_PRL__120_087201}. Neither magnetic ordering nor spin freezing is observed with the temperature down to 0.26 K evident by spectral analysis and spin-lattice relaxation rate (SLRR) measurements. For the spin excitations, the SLRRs show a temperature-independent behavior for 0.26 K$<T<$ 10 K under a weak magnetic field. With stronger magnetic fields, we observe a spin excitation gap with the gap size proportional to the field intensity. These observations imply an unusual quantum disordered ground state with strong spin fluctuations.

Both polycrystals and single crystals of Yb(BaBO$_{3}$)$_{3}$ are used in this work. The polycrystals are synthesized by the solid reaction method\cite{Gao_JAC_745_396} and the single crystals are obtained by the conventional flux method. For the NMR study, about ten milligram of polycrystals and single crystals with typical dimensions of $4\times4\times 0.1$ mm$^3$ are selected. The single crystal is placed on a piezoelectric nano-rotation stage for precise alignment of the field direction. Our NMR measurements are conducted on the $^{11}$B nuclei($\gamma_n=13.655$ MHz/T, $I=3/2$) with a phase-coherent NMR spectrometer. The spectrum is obtained by summing up or integrating the spin-echo intensities at different frequencies or magnetic fields. The SLRR is measured by the standard inversion-recovery method.

\begin{figure}
\includegraphics[width=7cm, height=6cm]{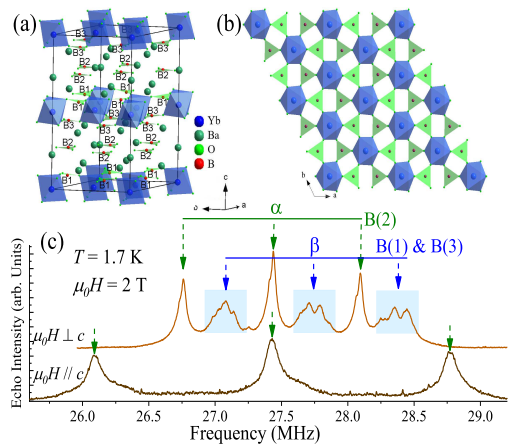}
\caption{\label{struc1}(color online) (a) A sketch of the crystalline structure of Yb(BaBO$_{3}$)$_{3}$. The inequivalent positions of $^{11}$B sites are denoted by "B1" to "B3". (b) Top view of the triangular arrangement of magnetic Yb$^{3+}$ cations. (c) Typical $^{11}$B NMR spectra with the magnetic field applied perpendicular or parallel to the $c$-axis of the single crystal.
}
\end{figure}

The Yb(BaBO$_{3}$)$_{3}$ crystalizes in the hexagonal structure with space group $P6_3cm$\cite{SM, Gao_JAC_745_396}[See Fig.\ref{struc1}(a)]. The YbO$_6$ octahedrons locate on the crystalline $ab$-plane, and are inter-connected by the corner-shared BO$_3$ triangles. The triangular planes formed by Yb$^{3+}$ ions are interleaved with three layers of nonmagnetic Ba/B-O polyhedrons stacked layer upon layer along c-axis, and can be treated as quasi-two dimensional frustrated antiferromagnets\cite{Li_PRL_115_167203,Guo_J3M_472_104,Guo_IC_58_3308,Guo_PRM_3_094404}.

Typical single crystal $^{11}$B NMR spectra are presented in Fig.\ref{struc1}(c) for both field directions. With a in-plane magnetic field, the spectrum is composed by two groups of peaks denoted by the down arrows and shadows. The $^{11}$B nucleus has a nuclear spin of $3/2$, thus we can expect three NMR transitions for the nucleus in a non-zero local electric field gradient (EFG). In Yb(BaBO$_{3}$)$_{3}$, there exist three inequivalent $^{11}$B sites, with the inter-layer and in-plane boron sites respectively denoted by B2 and B1/B3 in Fig.\ref{struc1}(a). The nuclei contributing to the sharp $\alpha$ peaks have a weaker hyperfine coupling with the magnetic layer, as evidenced by the nearly temperature independent Knight shift (the relative line shift with respect to the Larmor frequency, shown below), and a much slower spin-lattice relaxation as compared with the $\beta$ peaks. Hence, the $\alpha$ peaks can be assigned to the B2 site, and $\beta$ peaks are from the in-plane B1/B3 site. Under a field along the crystalline $c$-axis, the $\beta$ peaks from B1/B3 sites smear out and become undetectable as a result of too fast spin-spin relaxations, again confirming their strong coupling to the magnetic site. The nuclear quadruple resonance frequency $\nu_Q$ is calculated to be $\sim1.34$ MHz for B2 and $\sim1.31$ MHz for B1/B3 at $T=1.7$ K.

\begin{figure}
\includegraphics[width=7cm, height=8cm]{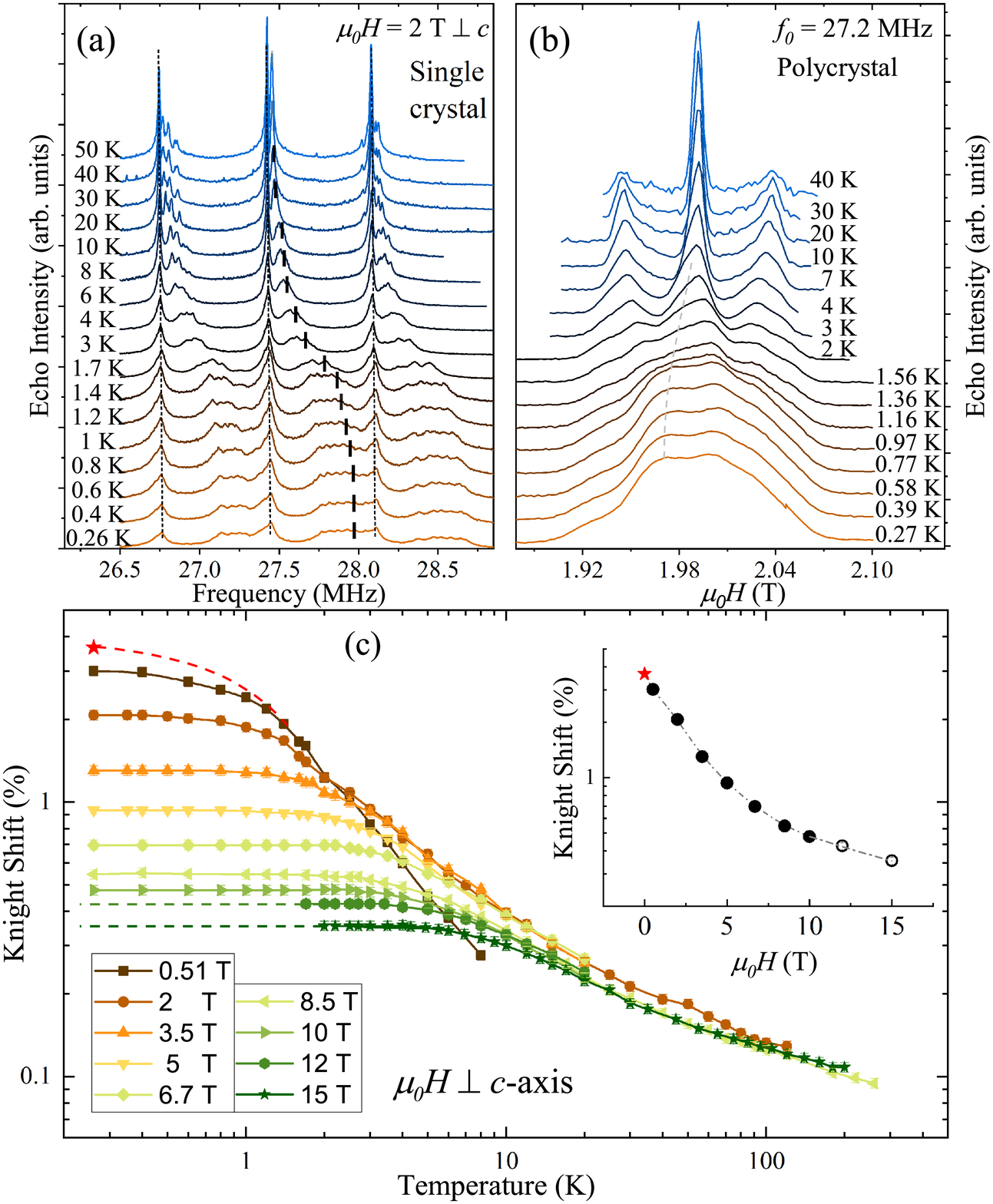}
\caption{\label{spec2}(color online)
(a) The frequency-swept $^{11}$B NMR spectra of the single crystal.
(b) The field-swept $^{11}$B NMR spectra of the polycrystal.
(c) The Knight shift with the magnetic field perpendicular to the crystalline $c$-axis as a function of temperature. The saturated Knight shift as a function of field intensity is shown in the inset.
}
\end{figure}

We present the spectra at different temperatures of the single crystal and  polycrystal respectively in Fig.\ref{spec2} (a) and (b). The two inequivalent B(1) and B(3) sites should give birth to double peaks at the high frequency side in the single crystal. However, we have observed multiple tiny peaks (This is more obvious in the high field spectrum ($\mu_0H=15$ T) at temperatures as high as $T=220$ K). This is more clear for the satellite peaks, which is much more sensitive to the structural inhomogeneity. The tiny splitting is supposed to result from the disorder effect.

Neither spin ordering nor freezing behavior is observed for the entire temperature region down to $T=0.26$ K. With the single crystal cooling down, both the central transition and satellites shift to a higher frequency. The frequency gap between the tiny peaks also increases, which is directly proportional to the Knight shift. For the polycrystal, the central peak also shift to the low field side, consistently indicating the enhanced Knight shift. In condensed matter NMR, there mainly exist two possible reasons for the line splitting or broadening, with different temperature and field dependence. One is the magnetically ordered state, where a non-zero hyperfine field contributed from the static ordered moment give rise to the line splitting (for the commensurate magnetic order) or typical double-horn like line shape (for the incommensurate order). The frequency gap can be viewed as the order parameter, and show weak field dependence as the hyperfine field results from the ordered moment instead of the electron spin magnetization. This is completely not applicable to the present case, as the frequency gap can be well scaled with the spin susceptibility reflected in Knight shift.

The other reason is the distributed Knight shift in the paramagnetic phase. In the ordinary case, the inhomogeneous Knight shift across the sample will result in obvious line broadening instead of the splitting. We think the $^{11}$B nuclei which are supposed to locate on B(1) and B(3) sites may spread to other inequivalent nearby sites as a result of the small ion radius, resulting to some different discrete hyperfine coupling constant. This should be the most possible reason for the line splitting observed in our sample.

To study the intrinsic spin susceptibility, we further plot the Knight shift as a function of temperature in Fig.\ref{spec2}(c) under different field intensities. The peak positions for the single crystal are determined by fits to the B(1) and B(3) central transitions with the multi-Lorentz function. The Knight shift is calculated based on the frequency of the peak at the high frequency side denoted by the black short lines in Fig.\ref{spec2}(a) with stronger hyperfine coupling to gain a better sensitivity, and make tracking of the peaks more easily. With the sample cooling down, all the Knight shifts share a similar temperature dependence, first increase mildly and begin to level off at low temperatures. Under high magnetic fields, the temperature for the level-off behavior appearing further rises and the saturated Knight shift is suppressed (see Fig.\ref{spec2}(c) inset). We further extrapolate the saturated Knight shift to zero field limit (shown by the $\star$ symbol), and its temperature dependence is proposed by the dashed line in Fig.\ref{spec2}(c).  The level-off behavior and its field dependence is related with the gapless spin excitations in the highly anisotropic spin system and their suppression under applied field, which will be further discussed in the last part.

Next, we study the novel spin excitations through the spin-lattice relaxation. The spin-lattice relaxation rate $(T_1)^{-1}$ formulated as $(T_1)^{-1}\propto T\sum_{\overrightarrow{q}}|A(\overrightarrow{q})|^2[\chi^{''}(\overrightarrow{q},\omega_L)]$, is a good probe for the low-energy spin fluctuations in solids, where $A(\overrightarrow{q})$ and $\chi^{''}(\overrightarrow{q},\omega_L)$ respectively denote the hyperfine coupling tensor as a function of the wave vector $\overrightarrow{q}$ and the imaginary part of the dynamic susceptibility at the nuclear Larmor frequency $\omega_L$.

\begin{figure}
\includegraphics[width=7cm, height=6.5cm]{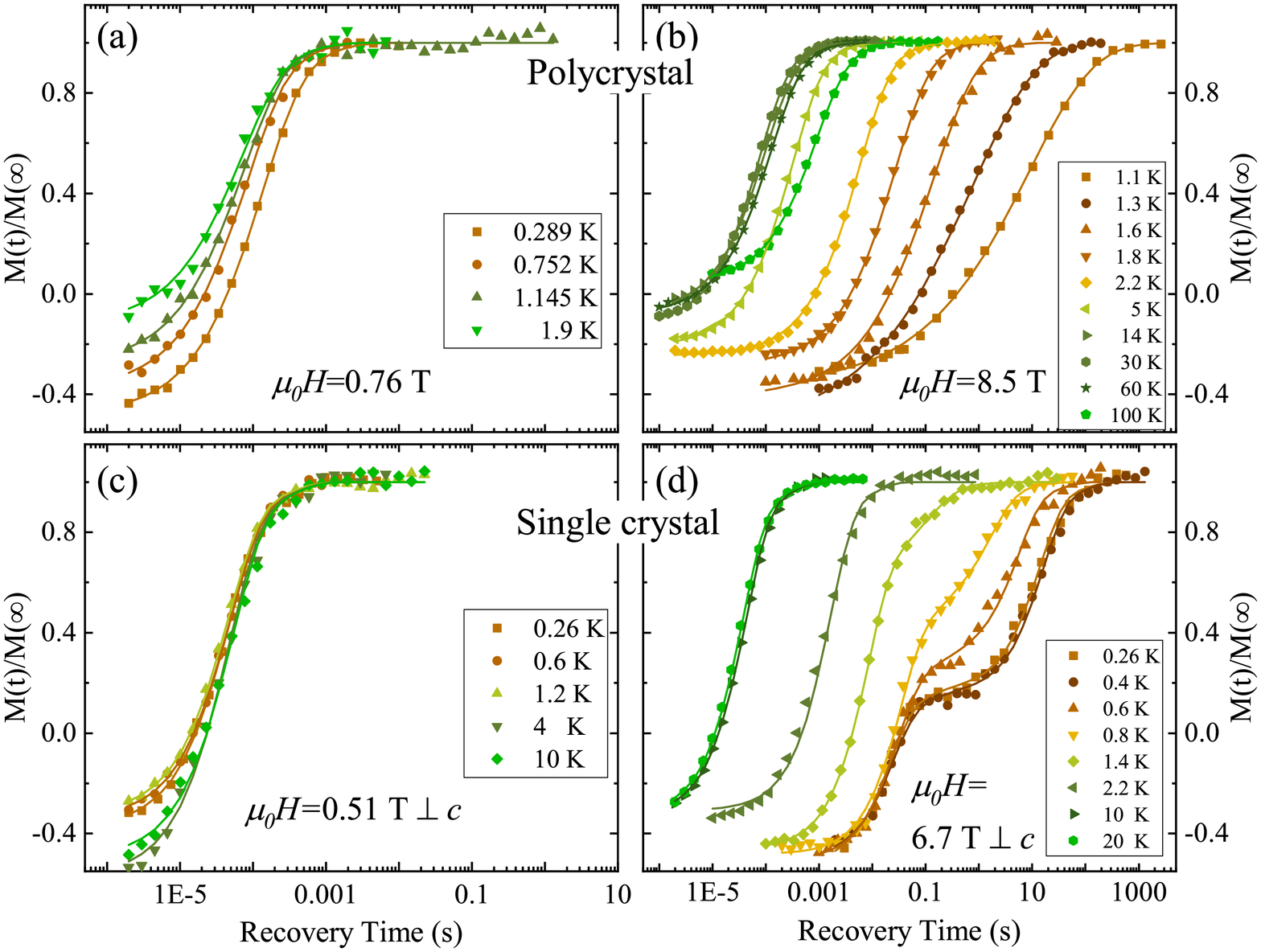}
\caption{\label{recovery3}(color online)
  Typical nuclear magnetization recovery curves for the polycrystal ((a) and (b)) and the single crystal ((c) and (d)). The solid lines are fittings to the functional forms described in the main text.
}
\end{figure}

\begin{figure}
\includegraphics[width=7cm, height=9cm]{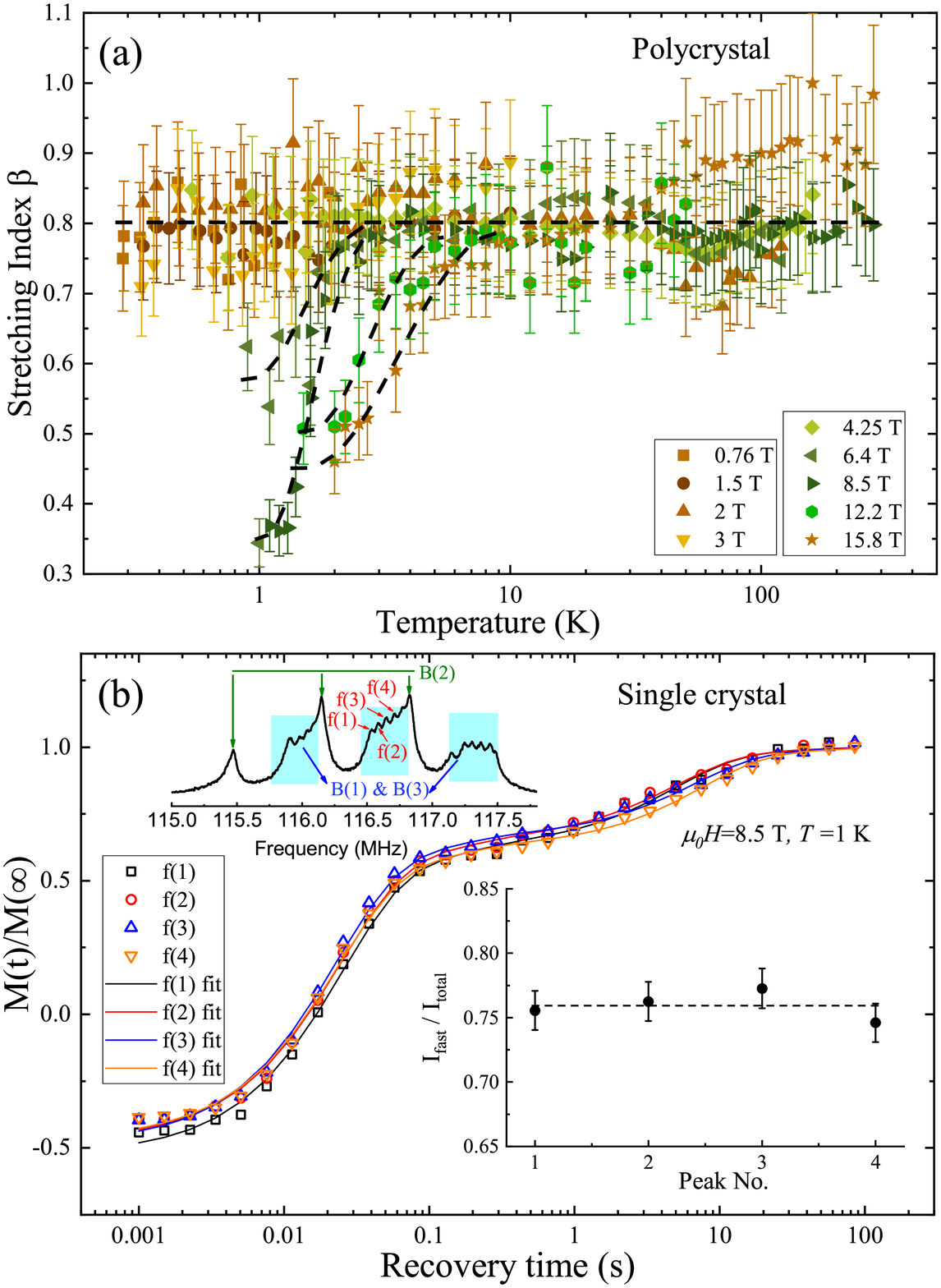}
\caption{\label{stretch4}(color online)
  (a) The temperature dependence of the stretching index $\beta$ of the polycrystal under different field intensities.
  (b) The nuclear magnetization recovery of the single crystal at different frequencies of the B(1) and B(3) spectra. The f(1) to f(4) respectively mark the different minor peaks contributed by the B(1) and B(3) sites, where the measurements of the SLRR are performed (left inset). The spectral weight of different $T_1$-components is shown in the right inset.
}
\end{figure}

The SLRR are measured by the standard inversion-recovery method with a better accuracy, and determined by fitting the observed time dependent nuclear magnetization to the recovery function. We show some representative spin recovery curves in Fig.\ref{recovery3}. For the polycrystal (Fig.\ref{recovery3} (a) and (b)), The stretching behavior in the recovery curve is always observed due to the overlap of the resonance peaks resulting from line broadening and also the distribution of SLRR values. The temperature dependence of the index $\beta$ is further shown in Fig.\ref{stretch4} (a) for different field intensities. For the low field region ($\mu_0H<=4.25$ T) and high field ($\mu_0H>=6.4$ T) high temperature region, the stretching index is around 0.8. With $\mu_0H>=6.4$ T, the index drops gradually at sufficient low temperatures.

For the single crystal (Fig.\ref{recovery3} (c) and (d)), the standard relaxation function for nuclei with $I=3/2$ is used to fit the spin recovery for the low field region ($\mu_0H<=3.5$ T) and high field high temperature region ($\mu_0H>3.5$ Tesla and $T>=$1.7 K), where only one $T_1$-component is observed. For the high field low temperature region, the function with double $T_1$-component is used to fit the recovery curve. The stretching behavior seen in the polycrystal is not present here. We have carefully checked the frequency dependence of the SLRRs, and show the representative nuclear magnetization recovery curves at different frequencies of the central transition of B(1) and B(3) sites in Fig.\ref{stretch4} (b). Obviously, the relaxation behaviors are very similar to each other. The variation of the relative weights between two different components is about 3.5\% and the difference of the SLRR is within 15\%, indicating the uniform spin excitations in the single crystal.

\begin{figure}
\includegraphics[width=7cm, height=8cm]{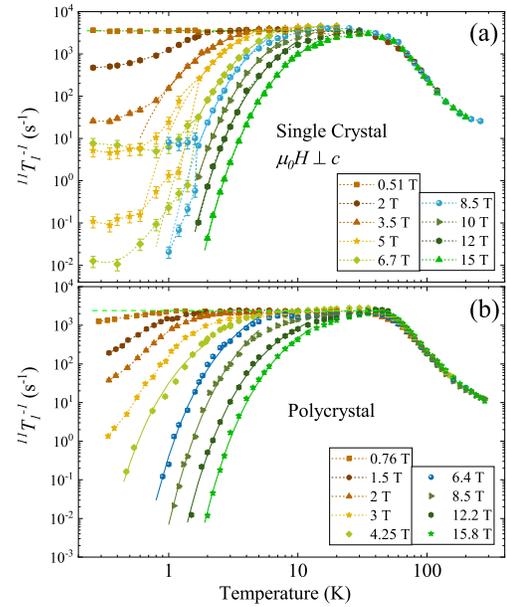}
\caption{\label{slrr5}(color online)
  The temperature dependence of the SLRR ($(^{11}T_1)^{-1}$) for $\mu_0H\perp c-$axis in the single crystal (a) and in the polycrystal (b). For the field higher than 5 Tesla, double $T_1$ components are needed to fit the nuclear magnetization recovery curve below $T\sim 1.7$ K, and both the $(^{11}T_1)^{-1}$ values are also shown in (a). We think the component with faster relaxation rates is contributed from the disorder effect, corresponding to the in-gap state observed in neutron scattering\cite{Paddison_NP_13_117}.
  The solid lines are fits to the thermally activation behavior (See the text).
  The $^{11}T_1^{-1}$ measurement is made at the frequencies corresponding to the B1/B3 sites in the single crystal.
}
\end{figure}

In Fig.\ref{slrr5}, we present the temperature dependence of $(^{11}T_1)^{-1}$ under a wide field intensity range to discuss the spin excitations in Yb(BaBO$_{3}$)$_{3}$. For the single crystal, the $(^{11}T_1)^{-1}$ is measured at the same peak, whose Knight shift is shown in Fig.\ref{spec2}(c). With the sample cooling from room temperature to $T\sim70$ K, the $(^{11}T_1)^{-1}$ strongly increases, and shows a field-independent characteristic. Below $T\sim70$ K, the $(^{11}T_1)^{-1}$ begins to flatten out, and shows a precise temperature-independent behavior below $T=10$ K down to 0.26 K under a in-plane 0.51 Tesla weak magnetic field. With strengthening the magnetic field intensity, the $(^{11}T_1)^{-1}$ drops gradually at low temperatures, and shows a fan-like shape. For both our single crystals and polycrystals, very similar behavior are observed (See Fig.\ref{slrr5} (a) and (b)), confirming these phenomena to be completely intrinsic.

The $(^{11}T_1)^{-1}(T)$ behavior for the high temperature region (70 K$<T<300$ K) results from the slowing down of spin fluctuations contributed by the thermally excited Yb$^{3+}$ multiplet with strong correlations. This is consistent with the previous reported large $|\theta_w|$ obtained from the high temperature $dc$ susceptibility\cite{Gao_JAC_745_396}. More intriguing is the quantum spin excitations reflected in the low temperature $(^{11}T_1)^{-1}$ behavior.

Neither $\lambda-$like critical slowing down behavior typical for spin ordering nor "hump"-like slow spin dynamics resulting from spin freezing\cite{Shockley_PRL_115_047201} is observed in our sample, again demonstrating the absence of magnetic ordering or freezing with the temperature down to $T\sim0.26$ K, and further showing the strong magnetic frustration effect. The temperature independent SLRR directly indicate the maintaining spin excitations down to the temperature far below $|\theta_w|\sim2.3$ K, and demonstrate a constant sum of dynamic spin susceptibility in the momentum space.

\begin{figure}
\includegraphics[width=7cm, height=9cm]{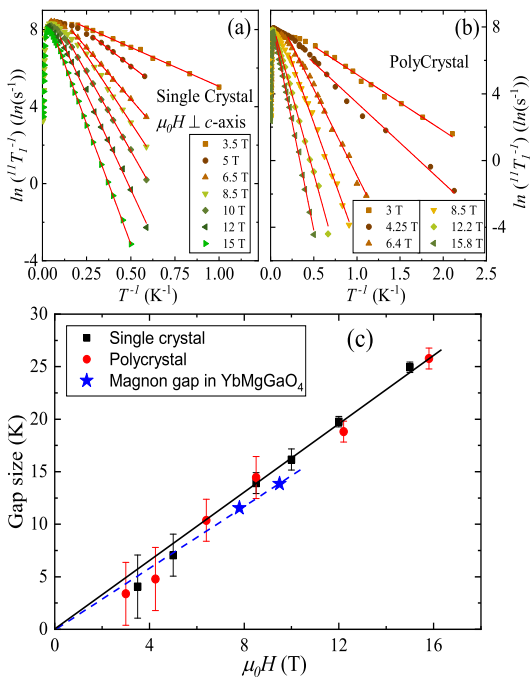}
\caption{\label{gap6}(color online)
  the natural logarithm of the SLRR ($\ln((^{11}T_1)^{-1})$) as a function of the reciprocal of temperature ($T^{-1}$) for the single crystal (a) and polycrystal (b) samples. The solid red lines are linear fits to the data. (c) The field dependence of the spin excitation gap obtained from the above fittings. The field dependence of the spin-wave excitation gap in YbMgGaO$_4$\cite{Shen_NC_9_4138,Paddison_NP_13_117} is also shown.
}
\end{figure}

We step further to explore the properties of novel quantum excitation by tracking its evolution under different magnetic field. In Fig.\ref{gap6}(a) and (b), we plot the $\ln((^{11}T_1)^{-1})$ versus $T^{-1}$, and fit the data with a linear function to demonstrate the magnetic field induced spin excitation gap. The gap size obtained from the fittings is shown in Fig.\ref{gap6}(c) as a function of the field intensity. Surprisingly, the spin excitation gap size is simply proportional to the applied field intensity.

The spin excitation behavior seen in this frustrated triangular lattice compound Yb(BaBO$_3$)$_3$ is very unusual. First, at the low field side, $(^{11}T_1)^{-1}(T)$ shows a precise temperature-independent behavior at low temperatures, and the Knight shift also show a level-off behavior.
Actually, this behavior is reported previously in YbMgGaO$_4$\cite{Li_PRL__117_097201}, NaYbS$_2$\cite{Sarkar_arxiv}, NaYbO$_2$\cite{Ding_PRB_100_144432} as well as other geometrically frustrated magnets\cite{Uemura_PRL_73_3306,Clark_PRL_110_207208,Fak_PRL_109_037208,Gomilsek_PRB_93_060405}.
The SLRR directly reflects the low energy spin excitations summarized in the momentum space, while the the Knight shift mainly measures the spin susceptibility at $\overrightarrow{q}=0$. The level-off behavior suggests the strong spin excitations maintaining at low temperatures
have a very weak $\overrightarrow{q}$ dependence. This fact is consistent with the proposed U(1) QSL ground state with spinon fermion surface\cite{Motrunich_PRB_72_045105,Lee_PRL_95_036403,Lee_PRB_46_5621}. Further identifications by measuring the full spin excitation spectrum is needed.

Second, at the high field side, a field induced spin excitation gap is observed, with the gap size proportional to the field intensity. We think that this gap must not be the spinon excitation gap under magnetic field, even if there exist spinon excitations at the low field side. Recent theoretical study suggests that the spinon bands splits in magnetic field due to the Zeeman effect, and the energy gap is directly proportional to the field intensity\cite{Li_PRB_96_075105}. However, this theory is based on the fact that the spinon remains a good description of the spin excitation in the weak field regime\cite{Shen_NC_9_4138}. The well-defined dispersive spin-wave excitation is already observed by the inelastic neutron scattering for fields above $\mu_0H=7.8$ T in YbMgGaO$_4$\cite{Paddison_NP_13_117}. For the present sample, the saturated magnetic field is comparable to that in YbMgGaO$_4$ as seen from the susceptibility data\cite{Gao_JAC_745_396}. The QSL state should not be preserved for such a high magnetic field, and this field dependence of the gap size must have no obvious connections with the spinon excitations. Thus, the most possible origin of the evolution of the field induced spin gap should be related with the spin-wave excitations at the strong magnetic field regime\cite{Li_PRB_94_035107}. The suppressed Knight shift at the high field side is also related with the the gapped out low-energy spin excitations\cite{Paddison_NP_13_117,Shen_NC_9_4138}.

At last, we would like to compare our sample with the recently discovered Yb-based 112 system with the chemical formula AYbX$_2$ (A=Na, Cs; B=O, Se)\cite{Zhang_CPL_35_117501}.
For field up to $\mu_0H=15.8$ T, typical behaviors for the spin ordering or freezing are completely absent in our samples. This indicates that the spin system stays far away from the magnetic instable point, in sharp contrast with that in the recently discovered $J_{eff}=1/2$ triangular lattice NaYbO$_2$\cite{Ranjith_PRB_99_180401,Bordelon_NP_15_1058}, NaYbSe$_2$\cite{Ranjith_arxiv_1911_12712}, CsYbSe$_2$\cite{Xing_arxiv_1911_12286} and the Kitaev QSL candidate material $\alpha$-RuCl$_3$\cite{Baek_PRL_119_037201,Zheng_PRL_119_227208}. Thus, from this aspect, our sample has supplied a better reference compound for YbMgGaO$_4$.

To conclude, we performed the first NMR study of the novel spin excitations in the newly discovered $J_{eff}=1/2$ triangular lattice compound Yb(BaBO$_3$)$_3$ . Absence of any spin ordering or freezing is evidenced by both the spectral analysis and the SLRR measurements, with the temperature down to 0.26 K. At low magnetic fields, The temperature independent $(^{11}T_1)^{-1}$ at low temperatures indicates the maintaining of strong quantum spin excitations and a constant sum of dynamic spin susceptibility in the momentum space. These observations indicate the strong magnetic frustration effect, and a novel quantum ground state with strong spin fluctuations. Under high magnetic fields, the field-induced spin excitation gap is observed, with the gap size proportional to the field intensity. We think this gap should be the gap of the dispersive spin-wave excitations at the strong magnetic field regime of the QSL state.

We thank Y. S. Li, Y. P. Cai, J. H. Zhou and Gang Chen for helpful discussions. This research was supported by the National Key Research and Development Program of China (Grant No. 2016YFA0401802), the National Natural Science Foundation of China (Grants No. 11874057,11504377, 11574288, 11874158, U1732273 and 21927814) and the Users with Excellence Program of Hefei Science center CAS (Grant No. 2019HSC-UE008). A portion of this work was supported by the High Magnetic Field Laboratory of Anhui Province.


\end{document}